\documentclass[aps, notitlepage, nofootinbib, twocolumn]{revtex4-1}

\usepackage{geometry}
\usepackage{bm}

\begin{document}

\title{The Universe is not a Computer}
\author{Ken Wharton}
\affiliation{Department of Physics and Astronomy, San Jos\'{e} State University, San Jos\'{e}, CA 95192-0106}
\begin{abstract}

\setlength{\baselineskip}{1.1\baselineskip} 

When we want to predict the future, we compute it from what we know about the present.  Specifically, we take a mathematical representation of observed reality, plug it into some dynamical equations, and then map the time-evolved result back to real-world predictions.  But while this computational process can tell us what we want to know, we have taken this procedure too literally, implicitly assuming that the universe must compute itself in the same manner.  Physical theories that do not follow this computational framework are deemed illogical, right from the start.  But this anthropocentric assumption has steered our physical models into an impossible corner, primarily because of quantum phenomena.  Meanwhile, we have not been exploring other models in which the universe is not so limited.  In fact, some of these alternate models already have a well-established importance, but are thought to be mathematical tricks without physical significance.  This essay argues that only by dropping our assumption that the universe is a computer can we fully develop such models, explain quantum phenomena, and understand the workings of our universe. (This essay was awarded third prize in the 2012 FQXi essay contest; a new afterword compares and contrasts this essay with Robert Spekkens' first prize entry.)

\end{abstract}

\maketitle

\setlength{\baselineskip}{1.35\baselineskip} 

\section{Introduction: The Newtonian Schema}

Isaac Newton taught us some powerful and useful mathematics, dubbed it the ``System of the World", and ever since we've assumed that the universe actually runs according to Newton's overall scheme.  Even though the details have changed, we still basically hold that the universe is a computational mechanism that takes some initial state as an input and generates future states as an output.  Or as Seth Lloyd says, ``It's a scientific fact that the universe is a big computer".\cite{Lloyd}

Such a view is so pervasive that only recently has anyone bothered to give it a name: Lee Smolin now calls this style of mathematics the ``Newtonian Schema".\cite{Smolin}  Despite the classical-sounding title, this viewpoint is thought to encompass all of modern physics, including quantum theory.  This assumption that we live in a Newtonian Schema Universe (NSU) is so strong that many physicists can't even articulate what other type of universe might be conceptually possible.  

When examined critically, the NSU assumption is exactly the sort of anthropocentric argument that physicists usually shy away from.  It's basically the assumption that the way we humans solve physics problems must be the way the universe actually operates.  In the Newtonian Schema, we first map our knowledge of the physical world onto some mathematical state, then use dynamical laws to transform that state into a new state, and finally map the resulting (computed) state back onto the physical world.  This is useful mathematics, because it allows us humans to predict what we don't know (the future), from what we do know (the past).   But is it a good template for guiding our most fundamental physical theories?  Is the universe effectively a quantum computer?  This essay argues ``no'' on both counts; we have erred by assuming the universe must operate as some corporeal image of our calculations.

This is not to say there aren't good arguments for the NSU.  But it is the least-questioned (and most fundamental) assumptions that have the greatest potential to lead us astray.  When quantum experiments have thrown us non-classical curveballs, we have instinctively tried to find a different NSU to make sense of them.  Thanks to this deep bias, it's possible that we have missed the bigger picture: the mounting evidence that the fundamental rules that govern our universe cannot be expressed in terms of the Newtonian Schema.  It's evidence that we've so far found a way to fold back into an NSU, but at a terrible cost -- and without debate or recognition that we've already developed the core framework of a promising alternative.

Section II will detail the problems that arise when one tries to fit quantum phenomena into an NSU.  The following sections will then outline the alternative to the NSU and show how it naturally resolves these same problems.  The conclusion is that the best framework for our most fundamental theories is not the Newtonian Schema, but a different approach that has been developed over hundreds of years, with ever-growing importance to all branches of physics.  It seems astounding that we have not recognized this alternate mathematics as a valid Schema in its own right, but \emph{no} alternative makes sense if we've already accepted Lloyd's ``fact" that the universe is a (quantum) computer.  Only by recognizing that the NSU is indeed an \emph{assumption} can we undertake an objective search for the best description of our universe.    

\section{Challenges from the Quantum}

Until the 20th century, the evidence against the NSU was circumstantial at best.   One minor issue was that (fundamental) classical laws can equally well be run forward and backward -- say, to retrodict the historical locations of planets.  So there's nothing in the \emph{laws} to imply that the universe is a forward-running computer program, calculating the future from some special initial input.  Instead, every moment is just as special as every other moment. 

Of course, the same is true for a deterministic and reversible computer algorithm -- from the data at any time-step, one can deduce the data at all other time-steps.  Combined with a special feature of the Big Bang (its status as an ordered, low-entropy boundary condition), this concern mostly vanishes.\footnote{Although it does raise questions, such as \emph{why} the laws happen to be time-symmetric, if the boundary conditions are so time-asymmetric.}

But \emph{quantum} phenomena raise three major challenges to the NSU.  Standard quantum theory deals with each of them in basically the same way -- by assuming the NSU must be correct, and using suspiciously anthropocentric reasoning to recast the universe in an image of our quantum calculations.  

First, we have Heisenberg's uncertainty principle (HUP).  In the classical context of Heisenberg's original paper \cite{HUP}, this means we can never know the initial state of the universe with enough precision to compute the future.  This would not alone have challenged the NSU -- a universal computer could potentially use the full initial state, even if we did not know it.  But it weakens the above argument about how the Big Bang is special, because not even the Big Bang can beat the HUP -- as confirmed by telltale structure in the cosmological microwave background.  The special low-entropy order in the universe's initial state is accompanied by random, non-special, disorder. 

But conventional quantum theory rejects the above reading of the HUP.  In spirit with the NSU, the unknown quantities are no longer even thought to exist.  Note the implication: if we humans can't possibly know something, then the universe shouldn't know it either.  The Big Bang is restored as the universe's special  ``input'', and the NSU is saved.  But this step leads to new problems -- namely, we can't use classical laws anymore, because we don't have enough initial data to solve them.  To maintain an NSU, we're forced to drop down from classical second-order differential equations to a simpler first-order differential equation (the Schr\"odinger equation).  

This leads to the second major challenge -- the Schr\"odinger equation yields the wrong output.  Or more accurately, the future that it computes is not what we actually observe.  Instead, it merely allows us to (further) compute the probabilities of different possible outcomes.  This is a huge blow to the NSU.  Recall the three steps for the Newtonian Schema: 1) Map the physical world onto a mathematical state, 2) Mathematically evolve that state into a new state, and 3) Map the new state back onto the physical world.  If one insists on a universe that computes itself via the Schr\"odinger equation, the only way to salvage the NSU is to have step 3 be a probabilistic map.  (Even though the inverse of that map, step 1, somehow remains deterministic.)

Once again, since we are restricted from knowing the exact outcome, conventional quantum theory puts the same restrictions on the NSU itself.  In step 3, the story goes, not even the \emph{universe} knows which particular outcome will occur.  And yet one particular outcome \emph{does} occur, at least when one looks.   Even worse, the measurement process blurs together steps 2 and 3, affecting the state of the universe itself in a manner manifestly inconsistent with the Schr\"odinger equation.  The question of exactly where (and how) the universe stops using the Schr\"odinger equation is the infamous ``measurement problem" of quantum theory.  It becomes harder to think of the universe as computing itself if the dynamical laws are not objectively defined.

So it's perhaps unsurprising that many physicists imagine an NSU that ignores step 3 altogether; the universe is simply the computation of the ever-evolving Schr\"odinger equation, the mismatch with reality notwithstanding.  The only consistent way to deal with this mismatch is to take the Everettian view that our entire experience is just some small, subjective sliver of an ultimate objective reality -- a reality that we do \emph{not} experience.\cite{MWI}

Which brings us to the third challenge to the NSU: the dimensionality of the quantum state itself.  The phenomenon of quantum entanglement -- where the behaviors of distant particles are correlated in strikingly non-classical ways -- seems to require a quantum state that does not fit into the spacetime we experience.  The quantum state of a N-particle system formally lives in a ``configuration space" of 3N dimensions.  If the universe is the self-computation of such a state, then we live in a universe of enormous dimensionality.  Any consistent, NSU view of quantum theory (not merely the Everettians) must maintain that Einstein's carefully-constructed spacetime is fundamentally incorrect.  Instead, one must hold that Schr\"odinger accidentally stumbled onto the correct mathematical structure of the entire universe.

Of course, configuration space was not an invention of Schr\"odinger's; it continues to be used in statistical mechanics and other fields where one does not know the exact state of the system in question.  Poker probabilities, for example, are computed in such a space.  Only after the cards are turned face up does this configuration space of possibilities collapse into one actual reality.  

In the case of cards, though, it's clear that the underlying reality was there all along -- configuration space is used because the players lack information.  In the case of a theory that underlies everything, that's not an option.  Either the quantum state neglects some important ``hidden variables'', or else reality is actually a huge-dimensional space.  Conventional thinking denies any hidden variables, and therefore gives up on ordinary spacetime.  Again, note the anthropocentrism: we use configuration spaces to calculate entangled correlations, so the universe must \emph{be} a configuration space.\footnote{Like a poker player that denies any reality deeper than her own knowledge, imagining the face-down cards literally shifting identities as she gains more information.}  

The NSU becomes almost impossible to maintain in the face of all these challenges.  Treating the universe as a computer requires us to dramatically alter our dynamical equations, expand reality to an uncountable number of invisible dimensions, and finesse a profound mismatch between the ``output'' of the equations and what we actually observe. 

Of course, no one is particularly happy with this state of affairs, and there are many research programs that attempt to solve each of these problems.  But almost none of these programs are willing to throw out the deep NSU assumption that may be at ultimate fault.  This is all the more surprising given that there \emph{is} a well-established alternative to the Newtonian Schema; a highly regarded mathematical framework that is in many ways superior.  The barrier is that practically no one takes this mathematics literally, as an option for how the universe might ``work".  The next sections will outline this alternative and reconsider the above challenges.

\section{The Lagrangian Schema}

While a first-year college physics course is almost entirely dominated by the Newtonian Schema, some professors will include a brief mention of Fermat's Principle of least time.  It's a breathtakingly simple and powerful idea (and even pre-dates Newton's \emph{Principia}) -- it just doesn't happen to fit in with a typical engineering-physics curriculum.

Fermat's Principle is easy to state: Between any two points, light rays take the quickest path.  So, when a beam of light passes through different materials from point X to point Y, the path taken will be the fastest possible path, as compared to all other paths that go from X to Y.  In this view, the reason light bends at an air/water interface is not because of any algorithm-like chain of cause-and-effect, but rather because it's globally more efficient. 

However elegant this story, it's not aligned with the Newtonian Schema.  Instead of initial inputs (say, position and angle), Fermat's principle requires logical inputs that are both initial and final (the positions of X and Y).  The initial angle is no longer an input, it's a logical \emph{output}.  And instead of states that evolve in time, Fermat's principle is a comparison of entire paths -- paths that cannot evolve in time, as they already cover the entire timespan in question.  

This method of solving physics problems is not limited to light rays.  In the 18th century, Maupertuis, Euler, and Lagrange found ways to cast the rest of classical physics in terms of a more general minimization\footnote{Actually, extremization.} principle.  In general, the global quantity to be minimized is not the time, but the ``action''.  Like Fermat's Principle, this so-called Lagrangian Mechanics lies firmly outside the Newtonian Schema.  And as such, it comprises an alternate way to do physics -- fully deserving of the title ``Lagrangian Schema".

Like the Newtonian Schema, the Lagrangian Schema is a mathematical technique for solving physics problems.  In both schemas, one first makes a mathematical representation of physical reality, mapping events onto parameters.  On this count, the Lagrangian Schema is much more forgiving; one can generally choose any convenient parameterization without changing the subsequent rules.  And instead of a ``state", the key mathematical object is a scalar called the Lagrangian (or in the case of continuous classical fields, the Lagrangian density, $\cal{L}$), a function of those parameters and their local derivatives. 

There are two steps needed to extract physics from $\cal{L}$.  First, one partially constrains $\cal{L}$ on the boundary of some spacetime region (\emph{e.g.}, fixing X and Y in Fermat's Principle).  For continuous fields, one fixes continuous field parameters.  But only the boundary parameters are fixed; the intermediate parameters and the boundary derivatives all have many possible values at this stage.

The second step is to choose one of these possibilities (or assign them probabilistic weights).  This is done by summing the Lagrangian (densities) everywhere inside the boundary to yield a single number, the action $S$.  The classical solution is then found by minimizing the action; the quantum story is different, but it's still a rule that involves $S$.

To summarize the Lagrangian Schema, one sets up a (reversible) two-way map between physical events and mathematical parameters, partially constrains those parameters on some spacetime boundary \emph{at both the beginning and the end}, and then uses a global rule to find the values of the unconstrained parameters.  These calculated parameters can then be mapped back to physical reality.

\section{Newton vs. Lagrange}

There are two fairly-widespread attitudes when it comes to the Lagrangian Schema.  The first is that the above mathematics is just that -- mathematics -- with no physical significance.  Yes, it may be beautiful, it may be powerful, but it's not how our universe \emph{really} works.  It's just a useful trick we've discovered.  The second attitude, often held along with the first, is that action minimization is provably equivalent to the usual Newtonian Schema, so there's no point in trying to physically interpret the Lagrangian Schema in the first place.

To some extent, these two attitudes are at odds with each other.  If the two schemas are equivalent, then a physical interpretation of one should map to the other.  Still, the arguments for ``schema-equivalence'' need to be more carefully dismantled.  This is easiest in the quantum domain, but it's instructive to first consider a classical case, such as Fermat's Principle.

A typical argument for schema-equivalence is to use Fermat's principle to derive Snell's law of refraction, the corresponding Newtonian-style law.  In general, one can show that action minimization always implies such dynamic laws. (In this context, the laws are generally known as the Euler-Lagrange equations.)  But a dynamical law is not the whole Newtonian Schema -- it's merely step 2 of a three-step process.  And the input and output steps differ: Snell's law takes different inputs than Fermat's Principle and yields an output (the final ray position) that was \emph{already constrained} in the action minimization.  Deriving Newtonian results from a Lagrangian premise therefore requires a bit of circular logic.

Another way to frame the issue is to take a known embodiment of the Newtonian Schema -- a computer algorithm -- and set it to work solving Lagrangian-style problems with initial and final constraints.  The only robust algorithms for solving such problems are iterative\footnote{As in the Gerchberg-Saxton algorithm.\cite{GS}}, with the computer testing multiple histories, running back and forth in time.  And this sort of algorithm doesn't sound like a universe that computes itself -- the most obvious problem being the disconnect between algorithmic time and actual time, not to mention the infinite iterations needed to get an exact answer.

Still, conflating these two schemas in the classical domain where they have some modest connection is missing the point: These are still two different ways to solve problems.  And when \emph{new} problems come around, different schemas suggest different approaches.  Tackling every new problem in an NSU will therefore miss promising alternatives.   This is of particular concern in quantum theory, where the connection between the two schemas gets even weaker.  Notably, in the Feynman path integral (FPI), the classical action is no longer minimized when calculating probabilities, so it's no longer valid to ``derive" the Euler-Lagrange equations using classical arguments.\footnote{It's only when one combines the quantum wave equations with the probabilistic Born rule that FPI probabilities are recovered; see the discussion of Eqn (1) in \cite{WMP}.}

So what should we make of the Lagrangian Schema formulations of quantum theory?  (Namely, the FPI and its relativistic extension, Lagrangian quantum field theory, LQFT.)  Feynman never found a physical interpretation of the FPI that didn't involve negative probabilities, and LQFT is basically ignored when it comes to interpretational questions.  Instead, most physicists just show these approaches yield the same results as the more-typical Newtonian Schema formulations, and turn to the latter for interpretational questions.  But this is making the same mistake, ignoring the differences in the inputs and outputs of these two schemas.  It's time to consider another approach: looking to the Lagrangian Schema not as equivalent mathematics, but as a different framework that can provide new insights. 

\section {Quantum Challenges in a Lagrangian Light}

Section II outlined three challenges from quantum theory, and the high cost of answering them in the NSU framework.  But what do these challenges imply for an \emph{LSU}?  How would the founders of quantum theory have met these challenges if they thought the universe ran according to the mathematics of the Lagrangian Schema -- not as a computer, but rather as a global four-dimensional problem that was solved ``all at once"?  Surprisingly, the quantum evidence hardly perturbs the LSU view at all.  

The first challenge was the uncertainty principle, but the classical LSU had this built in from the start, because it never relied on complete initial data in the first place.  Indeed, for classical particles and fields, there's a perfect match between the initial data one uses to constrain the Lagrangian and the amount of classical data one is permitted under the HUP.  In Fermat's principle, if you know the initial light ray position, the HUP says you \emph{can't} know the initial angle.  

Curiously, this ``perfect match" is only one-way.  The HUP allows more ways to specify the initial data than is seemingly permitted by the Lagrangian Schema.  For example, the HUP says that one can know the initial position \emph{or} the angle of the light ray, but Fermat's principle only works with constrained initial positions.  

But this is not a problem so much as a suggested research direction, evident only to a Lagrangian mindset.  Perhaps the HUP is telling us that we've been too restricted in the way we've fixed the initial and final boundaries on classical Lagrangians.  The natural question becomes: What would happen if we required action-minimization for \emph{any} HUP-compatible set of initial and final constraints?  For classical fields, the answer turns out to be that such constraints must be roughly quantized, matching equations that look like quantum theory.\cite{Wharton3}

Because the LSU doesn't need complete initial data to solve problems, there's nothing wrong with the second-order differential equations of classical physics (including general relativity, or GR).  With this change, one can revive Heisenberg's original interpretation of the HUP, yielding a natural set of initially-unknown ``hidden variables" (such as the ray angles in Fermat's Principle).  In simple cases \cite{KGE}, at least, these hidden variables can not only explain the probabilistic nature of the outcomes, but can actually be computed (in hindsight, after the final boundary becomes known).  Furthermore, there's no longer a compelling reason to drop to the first-order Hamiltonian equations, the standard Newtonian Schema version of quantum theory.  And since it's this leap from Lagrangian to Hamiltonian that introduces many of the deepest problems for quantum gravity (the ``problem of time'', etc.), there are good reasons to avoid it if at all possible.

The Lagrangian Schema also provides a nice perspective on the second challenge: the failure of Newtonian-style equations to yield specific, real-world outcomes (without further probabilistic manipulations).  Recall this was the most brutal challenge to the NSU itself, raising the still-unresolved measurement problem and breaking the symmetry between the past and future.  But the LSU doesn't utilize dynamical equations, so it dodges this problem as well.  The temporal outcome is not determined by an equation, it's imposed as an \emph{input} constraint on $\cal{L}$.  And because of the time-symmetric way in which the constraints are imposed, there's no longer any mathematical difference between the past and future; both constraints directly map to the real world, without further manipulation.  In fact, the Lagrangian procedure of ``fixing'' the future boundary looks remarkably like quantum measurements, providing a new perspective on the measurement problem.\cite{FQXI3}  

A common complaint at this point is that the above feature is a bug, in that it somehow makes the Lagrangian Schema unable to make predictions.  After all, what we usually want to know is the outcome $B$ given the input $A$, or at least the conditional probability $P(B_i|A)$ (the probability of some possible outcome $B_i$ given $A$).  But if one particular outcome (say, $B_1$) is itself an external constraint imposed on $\cal{L}$, a logical input rather than an output, then we can't solve the problem without knowing the temporal outcome.  Furthermore, since in this case $B_1$ is 100\% certain, the other possibilities ($B_2$, $B_3$, etc.) can never happen, contrary to quantum theory.

But like the NSU, this complaint conflates our useful calculations with objective reality.  In truth, any particular observed event does indeed have a single outcome, with after-the-fact 100\% certainty.  If we don't yet know that outcome, we can still \emph{imagine} fixing different outcome constraints $B_i$, and using $\cal{L}$ to compute an expected joint probability $P(A,B_i)$ for each possibility.  It's then a simple matter to normalize subject to some particular initial condition $A$ and generate the conditional probabilities $P(B_i|A)$.  These probabilities live in our heads until the actual outcome appears and show us what has been the case all along, at which point we update our incomplete knowledge.  This is basic Bayesian probability (see the above poker example), and many have noted that it is a more natural interpretation of the standard quantum ``collapse''.\cite{Spekkens, Harrigan}

Finally, consider the challenge of quantum entanglement.  The problem with the NSU mindset is that it demands an input state that can compute all possible outputs, even if we don't know what type of measurement will eventually be made.  In N-particle systems, the number of possible future measurements goes up exponentially with N.  Keeping track of *all* possible future measurements requires a state that lives in an enormous configuration space.  It simply doesn't ``fit" in the universe we observe, or in Einstein's GR.

But as we've seen, the NSU conflates the information we humans need to solve a problem and the data that must actually correspond to reality.  In any particular case, a vast portion of this traditional quantum state turns out to be needless -- it never gets mapped to reality and is erased by the so-called ``collapse".  That's because all possible measurements \emph{don't} occur; only the actual measurement occurs.  Once the future measurement choice is known, the joint probabilities take on familiar forms, with descriptions that have exact mathematical analogies to cases that \emph{do} fit in spacetime.\cite{WMP,EPW}

Which brings us to the key point:  If one wants to ``fit" quantum theory into the spacetime of GR, one \emph{must} use the Lagrangian Schema, solving the problem ``all at once".  Only then can the solution take into account the actual future measurement -- which, recall, is imposed as a boundary constraint on $\cal{L}$.  So an LSU-minded physicist, when encountering entanglement, would have no reason to add new dimensions.  The ``spooky" link between entangled particles would merely be joint correlations enforced by virtue of both particles contributing to the same global action.\cite{EPW}

When viewed from a Lagrangian Schema mindset, the transition from classical to quantum phenomena is not only less jarring, but is arguably a natural extension.  Sure, some things have to change -- perhaps extending the principle of action minimization \cite{Wharton3} -- but they're changes that only make sense in an LSU, with no NSU translation.  Classical physics provided a few cases where the two Schemas seemed to almost overlap, perhaps lulling us into a feeling that these two approaches must \emph{always} overlap.  But the fact that quantum phenomena are so incomprehensible in an NSU, and more natural in an LSU, should make us consider whether we've been using a deeply flawed assumption all along.

\section{Conclusions: Our Lagrangian Universe} 

The best reasons for taking the Lagrangian Schema seriously lie in quantum theory, but there are other reasons as well.  It's the cleanest formulation of general relativity, with the automatic parameter-independence that GR requires, and bypasses problematic questions such as how much initial data one needs to solve the Newtonian-style version.  The LSU blends time and space together just like GR, while the NSU has to grapple with a dynamic evolution that seems to single out time as ``special''.  The standard model of particle physics is not a set of dynamic equations, but is instead a Lagrangian density, with deep and important symmetries that are only evident in such a framework.  Even NSU-based cosmological mysteries, such as why causally-disconnected regions of the universe are so similar, no longer seem as problematic when viewed in an LSU light.

But from the computational perspective of the NSU, any description of an LSU seems baffling and unphysical.  When trying to make sense of the LSU, a NSU-minded physicist might ask a number of seemingly-tough questions.  \emph{Which past events cause the future boundary constraint?  How do objects in the universe ``know'' what future boundary they're supposed to meet?  Doesn't Bell's Theorem \cite{Bell} prove that quantum correlations can't be caused by past hidden variables?}  A close look reveals these questions are already biased -- they all implicitly assume that we live in an NSU.  But without the mentality that the past ``causes" the future by some algorithmic process, the above questions are no longer well-posed.  

Constructing a complete theory built upon the Lagrangian Schema is a vast project, one that has barely even begun.  The necessary first step, though, is to recognize that the NSU is an assumption, not a statement of fact.  Even then, it will be difficult to put such a deep bias behind us completely, to distinguish our most successful calculations from our most fundamental physical models.  But it also wasn't easy to fight other anthropocentric tendencies, and yet the Earth isn't the center of the universe, our sun is just one of many, there is no preferred frame of reference.  Now there's one last anthropocentric attitude that needs to go, the idea that the computations we perform are the same computations performed by the universe, the idea that the universe is as `in the dark' about the future as we are ourselves.

Laying this attitude to one side, at least temporarily, opens up a beautiful theoretical vista.  We can examine models that have no Newtonian Schema representation, and yet nicely incorporate quantum phenomena into our best understanding of spacetime.  We can treat the universe as a global, four-dimensional boundary-value problem, where each subset of the universe can be solved in exactly the same manner, with exactly the same rules.  Stories can be told about what happens between quantum measurements, and those very measurements can be enfolded in a bigger region, to simultaneously tell a bigger story.  And most importantly, such models will suggest further models, with alterations that only make sense in a Lagrangian framework -- perhaps a local constraint like $\cal{L}$$=\!\!\!0$, or treating the Euler-Lagrange equations as just an approximation to a fundamentally underdetermined problem.  

It is \emph{these} models, the balance of the evidence suggests, that have a chance of representing how our universe \emph{really} works.  Not as we humans solve problems, not as a computer, but as something far grander.\\

\section*{Afterword: \\ On Spekkens' Winning Essay}

``The Universe is not a Computer" was awarded third prize in the 2012 FQXi Essay Contest, ranked behind other excellent essays -- most notably the first prize winner \cite{Spekkens2}, written by the Perimeter Institute physicist Robert Spekkens.  Like the above essay, Spekkens zeroed in on how most physical theories are framed using a states-plus-dynamics Newtonian Schema, although with a different focus and conclusion.  With these essays now being presented in the same volume, this afterword is an opportunity to compare and constrast these two viewpoints.

Spekkens' essay begins by noting that physical theories generally are divided into ``dynamics" (laws that implement time evolution) and ``kinematics" (the space of physical states permitted by a theory).  These, of course, are the key components of any theory that falls under the Newtonian Schema, described above.  After noting how very different theories are framed in precisely this manner, Spekkens' essay states that physicists ``typically agree that any proposal must be described in these terms".  

Both of our essays are in general agreement that it is {\em this framing of physical theories}, in terms of dynamics+kinematics, that contains a widespread mistaken assumption -- but we are in disagreement as to the precise nature of the mistake.  Spekkens' essay makes the excellent point that seemingly-different theories (which postulate different kinematics and dynamics) can in fact be empirically indistinguishable when these two components are taken together.  I agree with Spekkens that two such theories should not be viewed as competitive explanations but rather as essentially identical.  (One possible lesson for theorists might be that they are proposing too many different theories, and can focus on a bare few.)

On the other hand, the above essay argues that theorists have been far too conservative in postulating different theories, in that they almost exclusively are couched in the Newtonian Schema.  One can make a case that framing theories in terms of kinematics + dynamics is more an instinctive habit than a well-thought-out ``agreement'', and that new Lagrangian-schema approaches are needed.  The mistake, in this view, is that the kinematics + dynamics framework is too restrictive, not too permissive.

These different conclusions are not mutually exclusive.  Take Spekkens' example of how classical physics can be couched in terms of forces and Newton's laws (on one hand) and Hamiltonian dynamics (on the other).  These two theories are indeed empirically indistinguishable, and should not be thought of as essentially different.  But they also both fall under the Newtonian Schema. It is notable that classical Lagrangian mechanics does not specify {\em any} dynamics, and therefore lies in a different category of theory altogether (a category unaddresed in Spekkens' essay).  In this sense, our essays are both making the case that Newtonian Schema theories are more similar than they might appear, and the above essay is making the additional case that Lagrangian Schema theories are different and under-explored.

One counterpoint to this claim might be to note that Lagrangian Schema theories can {\em also} be expressed in terms of dynamics + kinematics; namely, there are {\em no} dynamical laws, and the allowed kinematical ``states'' are merely four-dimensional {\em histories} that obey certain restrictions.  In other words, Lagrangian Schema theories are all kinematics, no dynamics.  

Might Spekkens' claim for empirical indistinguishability perhaps be extended to apply to Lagrangian Schema theories as well, showing them to all be essentially equivalent to a class of Newtonian Schema counterparts?  After all, classical Lagrangian mechanics is empirically equivalent to Newtonian mechanics (if leaving aside the input/output differences discussed above), and the probabilities generated by the Feynman path integral are empirically equivalent to the probabilities generated by the combination of the Schr\"odinger equation and the Born Rule \cite{WMP}.  Combined with the many inter-Newtonian-Schema examples in Spekkens' essay, this may make it seem like such an argument might be successfully developed.

But the essential differences between three-dimensional states governed by dynamics and four-dimensional ``histories" with no dynamics is far more dramatic than these examples imply.  Indeed, counter-examples have recently been published \cite{Price2,Wharton14} demonstrating simple Lagrangian Schema toy models with no dynamical counterpart whatsoever.  And far from being some unimportant curiosity, it is this precise style of model that most naturally maps to the very quantum phenomena that defy Newtonian Schema explanations. 

For example, consider the discussion concerning kinematical- and dynamical-locality in Spekkens' essay.  There, the point was that since fully-local Newtonian Schema accounts run afoul of the Bell inequalities, trying to rescue kinematical locality was essentially impossible:  Any such theory would necessarily have dynamical nonlocality, and would therefore always be empirically indistinguishable from a theory with kinematical nonlocality.  But in the case of the Lagrangian Schema, {\em there is no dynamics}, local, nonlocal, or otherwise.  The promise of rescuing kinematical locality (as discussed in section V) is now far more than just an empty redefinition of terms -- indeed, it is one of the primary motivations for pursuing Lagrangian Schema explanations in the first place.

So despite my general agreement with almost everything in Spekkens' winning essay, that essay is still framed in the Newtonian Schema mindset that is arguably a deep and mistaken assumption in its own right.  The claim in Spekkens' abstract that ``A change to the kinematics of a theory... can be compensated by a change to its dynamics without empirical consequence" is not always true when there are no dynamics in the original theory (as per the counter-examples in \cite{Price2,Wharton14}).  Still, since it does appear that this claim is true for Newtonian Schema theories, Spekkens' essay will hopefully help to focus the debate where it is needed: not between empirically indistinguishable Newtonian Schema explanations of quantum phenomena, but rather between dynamical and ``all at once'' explanatory accounts of our universe.

\onecolumngrid

\end{document}